\documentclass[twocolumn,showpacs,amsmath,amssymb,superscriptaddress]{revtex4}

\usepackage{epsfig,psfrag}
\usepackage{dcolumn}
\usepackage{bm}
\usepackage{graphicx}
\usepackage{color}
\newcommand{\beq}{\begin{equation}}
\newcommand{\eeq}{\end{equation}}
\newcommand{\beqr}{\begin{eqnarray}}
\newcommand{\eeqr}{\end{eqnarray}}

\newcommand{\va}{{\vec{a}}}
\newcommand{\vk}{{\vec{k}}}

\newcommand{\vE}{{\vec{E}}}

\newcommand{\vF}{{\vec{F}}}
\newcommand{\vG}{{\vec{G}}}

\newcommand{\sigmab}{\mbox{\boldmath $\sigma $}}

\newcommand{\vb}{{\vec{b}}}

\def\half{{1\over2}}

\def\eqa{\begin{eqnarray}}
\def\eea{\end{eqnarray}}

\begin{document}

\title{Quarter-Filled Honeycomb lattice with a quantized Hall conductance}
\author{Ganpathy Murthy}
\affiliation{Department of Physics and Astronomy,
University of Kentucky, Lexington KY 40506-0055}
\author{Efrat Shimshoni}
\affiliation{Bar-Ilan University, Ramat Gan, Israel}
\author{R. Shankar}
\affiliation{Yale University, New Haven CT 06520}
\author{H.A. Fertig}
\affiliation{Department of Physics, Indiana University, Bloomington, IN}
\date{\today}
\begin{abstract}
We study a generic two-dimensional hopping model on a honeycomb
lattice with strong spin-orbit coupling, {\it without the requirement
that the half-filled lattice be a Topological Insulator}. For
quarter-(or three-quarter) filling, we show that a state with a
quantized Hall conductances generically arises in the presence of a
Zeeman field of sufficient strength. We discuss the influence of
Hubbard interactions and argue that spontaneous ferromagnetism (which
breaks time-reversal) will occur, leading to a quantized anomalous
Hall effect.
\end{abstract}
\vskip 1cm 
\maketitle
Topological Insulators (TIs) are a new state of matter with a bulk gap
but protected edge/surface modes\cite{TI-reviews}. Quantum Hall
states\cite{QHE} are examples of time-reversal violating
TIs. Recently, time-reversal invariant TIs have been predicted\cite{TI-reviews} and
seen in experiments\cite{TI-expts}. They are characterized by a nontrivial $Z_2$
index in the bulk, and the presence of chiral edge/surface states
which are robust against localization due to static disorder.

One of the simplest models of two-dimensional TIs is the Kane-Mele
model\cite{Kane-Mele}, which is a tight-binding model on a honeycomb
lattice with hopping and spin-orbit interactions. A more realistic
model is given by a monolayer of the 3D TI Alkali Iridate $A_2IrO_3$
(where $A=Na,\ Li$), where the various tight-binding hopping
parameters are known from a fit to {\it ab initio}
calculations\cite{Shitade}. We will be working with a generic model on
a honeycomb lattice which has no symmetries other than lattice
translations, $2\pi/3$ rotations, and time-reversal.

Our goal is to construct an experimentally realizable system with a
quantized Anomalous Hall effect (AHE)\cite{AHE}. The AHE has
historically described the effect of magnetic order (spontaneous or
otherwise) on the Hall conductance.  We will describe a proposal for
observing a quantized Hall conductance in the absence of a
perpendicular magnetic flux which can be realized, for example, in a
monolayer of Alkali Iridate doped to $\frac{3}{4}$-filling, and
subject to a Zeeman field perpendicular to the monolayer. Previous
proposals in this direction\cite{QAHE-previous} have also either
explicitly violated time-reversal, or appealed to spontaneous magnetic
order in interacting TIs.  Our proposal does require strong spin-orbit
couplings but does not require the material to be a TI at
half-filling.
\begin{psfrags}
\begin{figure}
\includegraphics[scale=0.35]{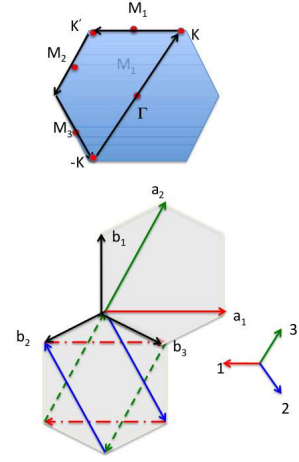}
\vspace*{-.1in}
\caption{Top: The Brillouin Zone of the honeycomb lattice. The $\Gamma$ and three $M$ points are time-reversal invariant, while the $K$ and $K'$ points go into each other under time-reversal. The arrows show the path along which the band structure will be plotted.
Bottom: The next-nearest neighbor hops in $A_2IrO_3$ are shown by
 dot-dashed red lines that go with $\sigma_1$, the solid blue that go
 with with $\sigma_2$, and the dashed green that go with
 $\sigma_3$. The primitive lattice vectors are $\va_1$ and $\va_2$
 while $\vb_1,\ \vb_2,\vb_3$ are nearest-neighbour hopping
 vectors. The auxiliary triad shows the projection of the original
 crystal axes on to the plane of the honeycomb lattice. Note that
 these directions match the nearest neighbor hops and the labels of
 the associated $\sigma$ matrices. }
\label{bz}
\end{figure}
\end{psfrags}

Generic models we consider share the following properties: They are
defined on a honeycomb lattice with two sites per unit cell. Including
spin, there are four bands. In the Brillouin Zone (BZ) shown in
Fig. (\ref{bz}) there are two kinds of special points. The $\Gamma$
and $M$ points are time-reversal invariant (modulo a reciprocal
lattice vector), while the $K$ and $K'$ points go into each other
under time-reversal. Time-reversal symmetry forces the bands to be
degenerate at the $\Gamma$ and $M_\alpha$ points. In a simple
spin-independent nearest-neighbor only tight-binding model (e.g.
Graphene) there are van Hove singularities at $\Gamma$ and $M_\alpha$,
whereas there are Dirac cones at $K$ and $K'$. Neither of these
features is generic under the addition of time-reversal invariant
spin-orbit couplings: The Dirac cones at the $K$ and $K'$ points can
be gapped by either a $L_zS_z$ type spin-orbit coupling, or by a
sublattice antisymmetric potential\cite{Kane-Mele}, while the van Hove
singularities are converted into Dirac cones by a Rashba
coupling. Fig. (\ref{bs-nozeeman}) shows such a generic time-reversal
invariant band structure.

We begin with a noninteracting model with three-fold rotational and
time-reversal symmetry, and focus on the generic Dirac crossings at
the $\Gamma$ and $M_{\alpha}$ points. Now consider a Zeeman field in
the $Z$ direction (perpendicular to the monolayer). This preserves the
three-fold rotational symmetry of the lattice, but breaks
time-reversal and will therefore provide a mass gap to the Dirac
points at $\Gamma$ and $M_\alpha$. We know that if the variation of a
parameter in the Hamiltonian leads to the touching of two bands at a
Dirac point for some critical value, a Chern number of $\pm1$ must be
exchanged between the two bands[8] as the parameter passes through
that value\cite{Dirac-exchange}. By the lattice rotational symmetry
all $M_\alpha$ must have the same exchange of Chern number. Since
there are three $M$ points and a single $\Gamma$ point, it is clear
that the Chern number exchange must be either $\pm2$ or $\pm4$. This
means that upon reversing the Zeeman coupling, the Chern number of the
lowest band must reverse, implying that its Chern number is either
$\pm1$ or $\pm2$. Of course, the Zeeman coupling should be strong
enough to create a hard gap (the bands should not overlap in energy)
in order to form an insulator. We will show that in a simple model of
$A_2IrO_3$ with Hubbard interactions, such a hard gap may form
spontaneously.

To demonstrate this behaviour in detail, we start with the
noninteracting hamiltonian for a freestanding monolayer of
Sodium/Lithium Iridate in the notation of \cite{Shitade}:
\begin{equation}
H_{SI}=-t\sum\limits_{\langle ij\rangle}\big(c^{\dagger}_ic_j+hc\big)+\sum\limits_{\langle\langle ij\rangle\rangle}c^{\dagger}_i{\hat t}_{ij} c_j
\label{hsi}\eeq
where the spin indices have been suppressed, $\langle ij\rangle$ is a
sum over nearest neighbors, $\langle\langle ij\rangle\rangle$ is a sum
over next-nearest neighbors, and the matrix ${\hat t}_{ij}$ in
spin-space is
\begin{equation}
{\hat t}_{ij}=t_0'+it'\sigma_a
\eeq
This hopping term is diagonal in the sublattice. Referring to
Fig. (\ref{bz}), the hopping is antisymmetric in the sublattice index
($A,\ B$). Further, each hop comes with a $\sigma_a$ matrix, the index
$a$ corresponding to the projection of the original crystal axes on to
the $111$ plane.  Thus, hopping along the primitive lattice vector
$\pm\va_1$ carries a $\sigma_1$, hopping along the vector
$\pm(\va_1-\va_2)$ carries a $\sigma_2$, and hopping along $\pm\va_2$
carries a $\sigma_3$. Let us denote the Pauli matrices in the
sublattice space by $\tau_a$. We obtain
\begin{equation}
h_{SI}(\vk)=t_0'F_0(\vk)-t\big(f_r(\vk)\tau_1-f_i(\vk)\tau_2\big)-t'\tau_3\big(\vF(\vk)\cdot\vec{\sigma}\big)
\eeq
where 
\beqr
f(\vk)=&f_r+if_i=e^{i\vk\cdot\vb_1}+e^{i\vk\cdot\vb_2}+e^{i\vk\cdot\vb_3}\\
F_0(\vk)=&\cos(\vk\cdot\va_1)+\cos(\vk\cdot\va_2)+\cos(\vk\cdot(\va_2-\va_1))\\
F_1(\vk)=&\sin(\vk\cdot\va_1)\\
F_2(\vk)=&\sin(\vk\cdot(\va_2-\va_1))\\
F_3(\vk)=&-\sin(\vk\cdot\va_2)
\eeqr

The parameters for the 111 plane of $Ir$ in $Na_2IrO_3$ are estimated
as $t=310K$, $t_0'=-130K$ and $t'=100K$.

One first diagonalizes $\vF\cdot\vec{\sigma}$, which has eigenvalues
$\pm|\vF|$, and obtains the energies at $\vk$ as
\beq
E_{\pm}(\vk)=t_0'F_0(\vk)\pm\sqrt{t^2|f|^2+t'^2|\vF|^2}
\eeq
It is seen that the four bands come in two pairs, degenerate at all
momenta. This is the result of inversion symmetry and clearly not
generic. For a monolayer on a substrate, a Rashba spin-orbit coupling
arising from the substrate $\vE$ field perpendicular to the 111 plane
is induced. A nearest-neighbor Rashba coupling leads to an additional
term
\beq
h_R(\vk)=i\lambda_R\vG(\vk)\cdot\sigmab\tau_+-i\lambda_R\vG^*(\vk)\cdot\sigmab\tau_-
\eeq
where
\beq
G_1(\vk)=\sqrt{\frac{2}{3}}\big(e^{i\vk\cdot\vb_1}-\half e^{i\vk\cdot\vb_2}-\half e^{i\vk\cdot\vb_3}\big)
\eeq
with cyclic permutations of $\vb_1,\ \vb_2,\ \vb_3$ defining $G_2,\
G_3$.  

Due to its special symmetries, this model does not show generic
behavior for $\lambda_R < t'$. For this parameter range, in addition
to the usual Dirac points at $\Gamma$, and $M_{\alpha}$ there are also
Dirac points at $K,K'$. For $t'> t'_{*}$ (where $t'_{*}$ is a function
of $\lambda_R$, but tends to $3t/8$ as $\lambda_R\to0$) the $M$ points
are unsplit. However, for $\lambda_R < t' < t'_{*}$, each $M$ point
splits into three Dirac points lying on the zone boundary. The central
point is still at $M$ and the other two ``satellites'' are
symmetrically distributed about it. At the special point
$\lambda_R=t'$, the satellites coalesce with the Dirac points at
$K,K'$. For $\lambda_R>t'$ the would-be Dirac points at $K,K'$ are
gapped, and the only Dirac points left are the generic ones. In the
following we will focus on this case.

A typical band structure including $\lambda_R=0.4t$ (satisfying
$\lambda_R>t'$) is shown in Fig. (\ref{bs-nozeeman}). One clearly sees
the Dirac-like crossings at the $\Gamma$ and $M_\alpha$ points.
\begin{psfrags}
\begin{figure}
\includegraphics[scale=0.35]{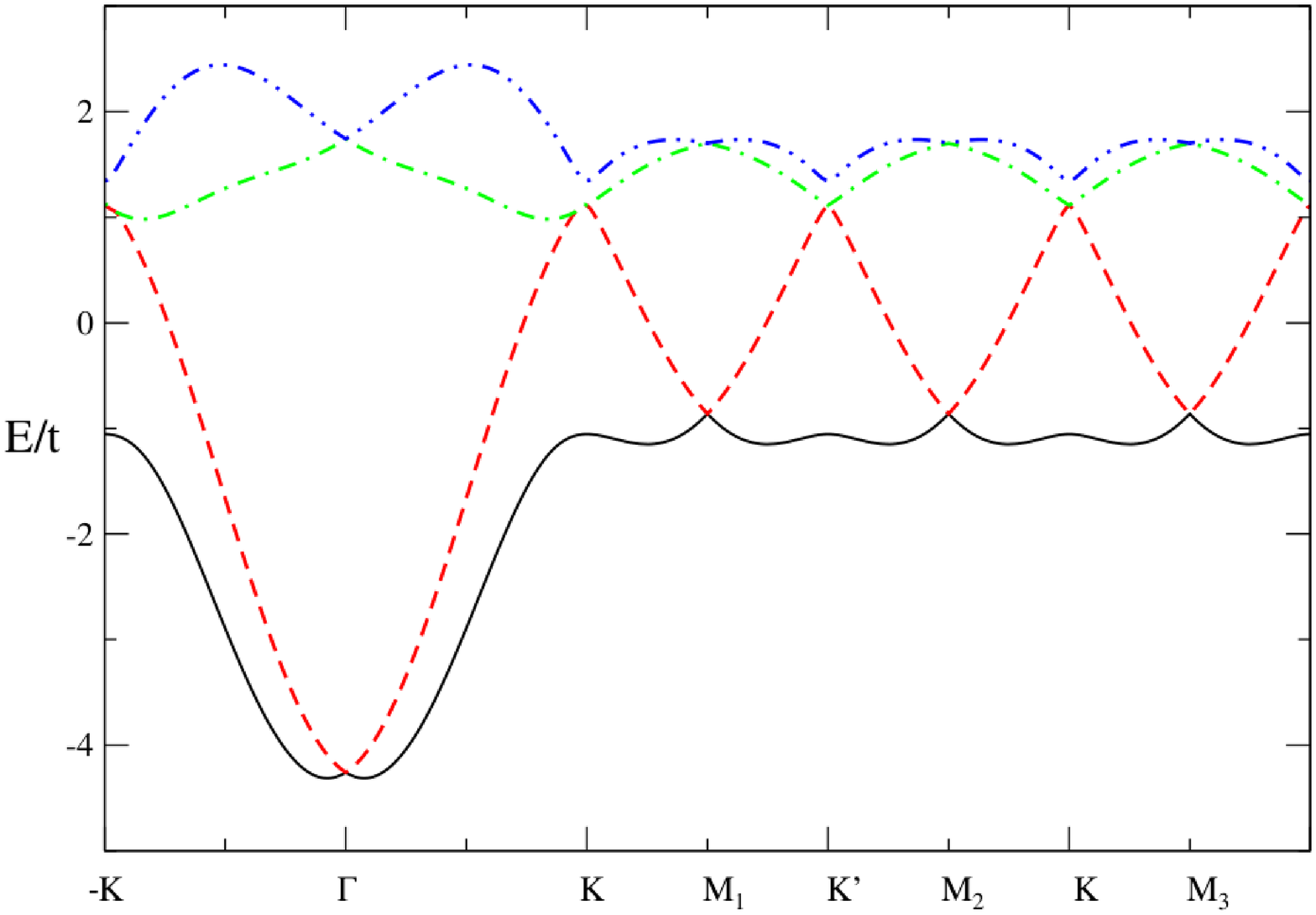}
\caption{The band structure for a monolayer of $A_2IrO_3$ with the 
addition of a Rashba term with $\lambda_R=0.4t$. The path in the BZ
followed is shown in Fig. (\ref{bz}).  The generic Dirac crossings at
the $\Gamma$ and $M_\alpha$ points can be seen. }
\label{bs-nozeeman}
\end{figure}
\end{psfrags}

For weak Zeeman coupling, a gap will open at the Dirac points, but the
two bands will overlap in energy (at different momenta), and the
system will be metallic at $\frac{1}{4}$ or
$\frac{3}{4}$-filling. Increasing the Zeeman coupling will lead to a
hard gap and thus a quantized Hall conductance. For the noninteracting
model, the critical Zeeman field is $0.37t$.  A calculation shows that
the Chern number for these parameters is $-1$. We note
parenthetically, that even for the nongeneric case $\lambda_R<t'$ the
Chern number of the highest energy band is nonzero ($1$), but
different from the above. Once a hard gap has been established, the
Chern number is robust against arbitrary deformations of the
hamiltonian which do not close the gap. In particular, one could
softly break the lattice rotational symmetry, add a Kane-Mele-type
spin-orbit coupling or a staggered sublattice potential, and also tilt
the Zeeman away from the perpendicular to the monolayer.

Let us now consider the influence of interactions. It is known that a
strong enough Hubard $U$ leads to an antiferromagnet in the
half-filled TI\cite{half-int-TI}. An on-site Hubbard interaction
(typically of size $\frac{U}{t}\simeq10$) spontaneously breaks
time-reversal symmetry at $\frac{3}{4}$ filling in the Hartree-Fock
approximation. Assuming only that lattice translation symmetry is
intact but allowing for the possibility of rotational and sublattice
symmetry breaking, we find in Hartree-Fock that the magnetization is
in the $Z$ direction (perpendicular to the monolayer), and the gap is
strongly enhanced. Because there is no spin-rotation symmetry in this
model, the $Z$ direction is the natural one, and the order-parameter
is Ising-like. This implies that the spin-waves are fully gapped, and
that there should be a finite-temperature phase transition even in the
two-dimensional material.

Of course, it could happen that the system chooses to break lattice
translation symmetry by forming a larger unit cell, as happens in
Graphene at a quarter-doping\cite{quarter-graphene}. However, in
Graphene, this is a result of Fermi-surface nesting, which is
sensitive to the details of the band structure, and is not generic
(there are also other competing states in Graphene at
$\frac{1}{4}$-filling). Furthermore, a strong enough Zeeman field in
the $Z$ direction will make our lattice symmetric solution
energetically favorable. We will leave a further discussion of the
issue of other possible ground states at quarter (or three-quarter)
filling\cite{other-gs-1/4} to future work. We now speculate on the
effects of long range (1/r) Coulomb interactions in this system, which
can coexist with the Hubbard interaction. Long-range Coulomb
interactions will be present because the material is insulating, and
will likely suppress Charge Density Wave states, which are one avenue
of translation symmetry breaking. Secondly, since the band has a
nonzero Chern number we expect skyrmionic excitations\cite{skyrmion}
to be the lowest energy charged excitations in some regime of
parameters. Finally, because the empty band is relatively flat, it may
represent an environment in which analogs of fractional quantum Hall
states will be stable when the band is partially
occupied\cite{flat-band,qi-TI-QH-map}.
\begin{psfrags}
\begin{figure}
\includegraphics[scale=0.35]{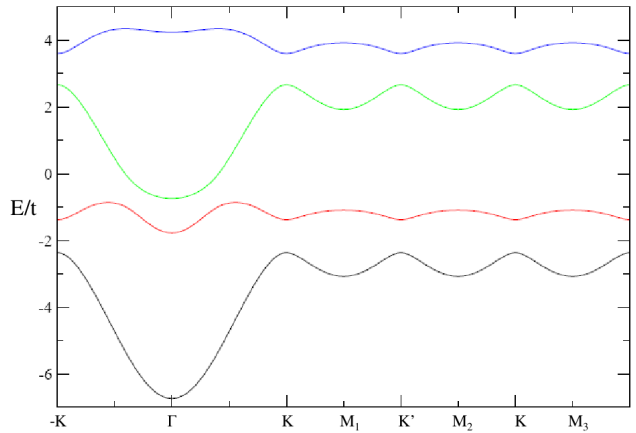}
\caption{The Hartree-Fock band structure at $U=10t$ and three-quarter filling in $A_2IrO_3$ with $\lambda_R=0.1t$. The path followed in the BZ is shown in Fig. (\ref{bz}). Note the hard gap and the flatness of the empty band (blue). }
\label{bsu10}
\end{figure}
\end{psfrags}

One can ask whether other simple lattices can support similar states
under generic conditions, where the only Dirac points are those
mandated by time-reversal symmetry. The simplest square lattice
two-band model does not support such states. The reason has to do with
the fact that the BZ has only two $M$ points. Repeating the argument
above Eq. (\ref{hsi}), we see that the Chern number exchange is $\pm1$
or $\pm3$, implying that the states have Chern number
$\pm\half,\ \pm\frac{3}{2}$, which is impossible for a 2D band
insulator. The half-filled triangular lattice with one band per unit
cell would have a tendency to become antiferromagnetic when
interacting. Thus the honeycomb lattice seems to be the simplest one
in which the desired properties can be realized.

In summary, we have shown that in a 2D honeycomb lattice at
$\frac{3}{4}$-filling, one can obtain a band with nonzero Chern number
with strong enough Rashba coupling, and an external Zeeman
coupling. Since the system does not have to be a TI at half-filling,
this increases the range of possible material realizations. One
promising material is $A_2IrO_3$ with $A=Na,\ Li$, with the Chern
number of the $\frac{3}{4}$-filled system being $\pm1$. In the bulk it
is half-filled and antiferromagnetically
ordered\cite{materials-issues} below $T_N=15K$.  A monolayer of the
$111$ $Ir$ plane sandwiched by two layers of $Na$ would induce
three-quarter doping\cite{qi-pvt}, where we expect it to become a
ferromagnetic quantized anomalous Hall insulator with an Ising-like
order parameter in the presence of Hubbard interactions of realistic
strength. It may be quite difficult to grow a single monolayer doped
in the manner described, but possibly the surface layer of a thin
enough film will display the same properties. We intend to investigate
this in future work.

We are grateful to the Aspen Center for Physics (NSF 1066293) for its
hospitality while this work was conceived and carried out. We would
also like to thank Greg Fiete, Yong-Baek Kim, Dung-Hai Lee, Karyn Le Hur, Steve
Simon, and particularly Xiaoliang Qi for illuminating
conversations. We are also grateful for partial support from
NSF-DMR-0703992 and NSF-PHY 0970069 (GM), ISF 599/10 (ES),
NSF-DMR-0901903 (RS), NSF-DMR-1005035 (HAF), and the US-Israel
Binational Science Foundation-2008256 (ES and HAF).

\end{document}